\documentstyle[preprint,epsf,aps]{revtex}

\begin{document}
\title{Traveling Granular Segregation Patterns in a Long Drum Mixer}
\author{Kiam Choo, T. C. A. Molteno and Stephen W. Morris}
\address{Department of Physics, University of Toronto,
Toronto, Ontario,
Canada, M5S 1A7}
\date{\today}
\maketitle

\begin{abstract}

Mixtures of granular media often exhibit size segregation along the axis of a partially-filled, horizontal, rotating cylinder. Previous experiments have observed axial bands of segregation that grow from concentration fluctuations and merge in a manner analogous to spinodal decomposition. We have observed that a new dynamical state precedes this effect in certain mixtures: bi-directional traveling waves. By preparing initial conditions, we found that the wave speed decreased with wavelength. Such waves appear to be inconsistent with simple PDE models which are first order in time. 

\end{abstract}

Many heterogeneous granular materials stubbornly refuse to mix when shaken or stirred.\cite{jaeger,bridge} Instead, the grains segregate by size. 
A simple apparatus, the horizontal drum 
mixer\cite{oyama,donald,dasgupta,savage,nakagawa,zik,hillpre,mri,mex}, dramatically illustrates this effect.  A binary mixture of grain sizes partially fills a horizontal cylinder, which is then rotated about its axis. The components separate into bands arranged along the long axis of the cylinder. In all experiments reported so far\cite{donald,dasgupta,nakagawa,zik,hillpre,mri,mex}, these bands show simple, irreversible merging dynamics eventually leading to a stable axial array, or to complete segregation.\cite{mex} This process has been likened to one-dimensional spinodal decomposition.\cite{zik} We report observations that demonstrate that axial segregation patterns can have a considerably richer dynamics than previously thought.  We find traveling waves of segregation that appear in a transient regime which precedes the formation of larger, nearly stationary bands.  The waves can travel in both directions and interpenetrate to form standing waves. This surprising phenomenon is difficult to reconcile with existing simple theories of segregation.\cite{savage,zik,hillpre}  

In fact, it is well-known that segregation occurs both radially and axially.\cite{donald,dasgupta,nakagawa,hillpre,mri}  In radial segregation, one component, usually the smaller one, builds up near the axis of rotation, forming a buried core.  This tendency is observed even in two-dimensional arrays of disks.\cite{2dexpt,2dtheory}   It is generally believed that radial segregation always precedes axial, although there is no clear understanding of why this is so. 

It has long been believed that axial segregation is driven by sorting processes operating at the flowing surface.\cite{donald,dasgupta,savage,nakagawa,zik,hillpre} The dynamic angle of repose, the angle at which the surface streams down for a given rotation frequency, is a function of the local composition of the mixture.  Once fairly complete segregation has occurred, one can easily observe a modulation of the dynamic angle of repose which is associated with the bands.  Furthermore, mixtures for which the dynamic angles become equal at some value of the rotation frequency exhibit remixing at that frequency.\cite{hillpre}  These observations have lead to models of surface-slope driven axial segregation which simply ignore the radially segregated core.\cite{savage,zik}  If these ideas are correct, one can learn all there is to know about axial segregation from observations of the streaming surface made from outside the cylinder.  This is the approach we have taken. Recent MRI measurements have cast doubt on this assumption, however.\cite{mri}  It was found that the axial bands were connected to the radially segregated core and indeed that they might be better viewed as undulations in the thickness of the core which break through the surface. This point of view was strengthened by the observation of subsurface bands.\cite{mri}  The modulation of the dynamic angle of repose would then be viewed as an effect of the segregation, rather than its cause.

The traveling waves are qualitatively different from the large, nearly stationary bands previously observed.  Although they are inferred from surface measurements, their most intriguing property, that they travel, is interesting regardless of the details of their subsurface structure, if any.  Our work leaves open the question of how the waves fit into the above contrasting points of view about the roles of axial {\it vs.} radial segregation. 

We considered how the waves might be explained in terms of simple one-dimensional segregation models based on the continuity equation.  The fact that the waves interpenetrate to form standing waves implies that they are nearly linear, at least at early times.  This feature is impossible to explain with simple PDE models which are first order in time.  Second order dynamics might arise if the segregation is coupled to collective oscillation modes of the grains\cite{fauve}, or to electrostatic effects.\cite{electro}  We have no evidence that either of these are present, although it is difficult to rule out the latter.  An important feature of the phenomenon is that the waves are only found in certain mixtures which have a greater volume fraction the larger component.  Most previous studies used equal mixtures, so they may have overlooked the effect.

The drum mixer consisted of a pyrex tube 109 cm long and 2.7 cm in diameter and was bounded by teflon covered plugs.  This long design was chosen to minimize end effects and to allow room for many bands to develop.  The tube was carefully loaded with a mixture of black sand\cite{hobbysand} and white table salt to a filled volume fraction of $0.28 \pm 0.02$. The components were sieved so that the black sand had sizes in the range $45 - 250 \mu$m, while the salt was $300 - 850 \mu$m. These sizes are similar to those used by Das Gupta {\it et al}. \cite{dasgupta} We define the size composition fraction $\phi$ of a mixture to be the volume of the white salt component divided by the sum of the volumes of the sand and salt components, where all volumes were measured before mixing.  Because of packing effects, the volume of a sand/salt mixture is smaller than the sum of their separate volumes.  Thus, $\phi$ is not exactly equivalent to the volume concentration of the salt in any homogeneous mixture, but is experimentally well-defined. We monitored the humidity during the experiment, and it was kept in the range 21-34\%.  Humidity has a bearing on possible electrostatic effects, as discussed below.  For all of the measurements we report here, the filled volume fraction of the tube and the angular frequency of rotation $\omega$ were kept fixed, while $\phi$ was varied over the range 0.33 - 0.79.  We used $\omega = 4.841 \pm 0.007 s^{-1}$, a frequency at which the mixture streams smoothly down its free surface.  

A CCD camera was positioned perpendicular to the streaming surface, which was uniformly lit by a single fluorescent tube. Images of the entire surface were digitized and time averaged over several frames. After division by a reference image and a correction for the camera lens distortion, these were spatially averaged across their height to produce one-dimensional snapshots of the surface color along the long axis of the mixer, which we define to be the $z$ axis. The surface color is a measure of the local concentration at the flowing surface. Obviously, this technique is insensitive to the concentration below the surface. 

Since this mixture segregates under almost any agitation, special care is required to load the tube with a premixed sample.  This was done by filling a U-shaped channel as long as the tube with the desired composition $\phi$.  The channel was then inserted lengthwise into the horizontal tube and inverted to deposit its contents.  This was done twice to load the tube. To prepare presegregated initial conditions, the channel was fitted with thin removable partitions so that the two components could be loaded separately into alternating sections of the desired lengths.  In this case, the composition fraction $\phi$ was determined by sieving after a run.

When the tube is loaded with a premixed sample with $\phi = 0.55$, we find that axial bands develop after a few minutes, which corresponds to a few hundred  rotations. No traveling waves are found. After some merging events, 16 - 20 stable black bands remain.  We have not systematically studied the long time evolution of these bands but it is evident that any subsequent merging must be on time scales of many hours.  These observations are similar to previous studies.\cite{donald,dasgupta,savage,nakagawa,zik,hillpre,mex} 

\begin{figure}
\epsfxsize =5in
\centerline{\epsffile{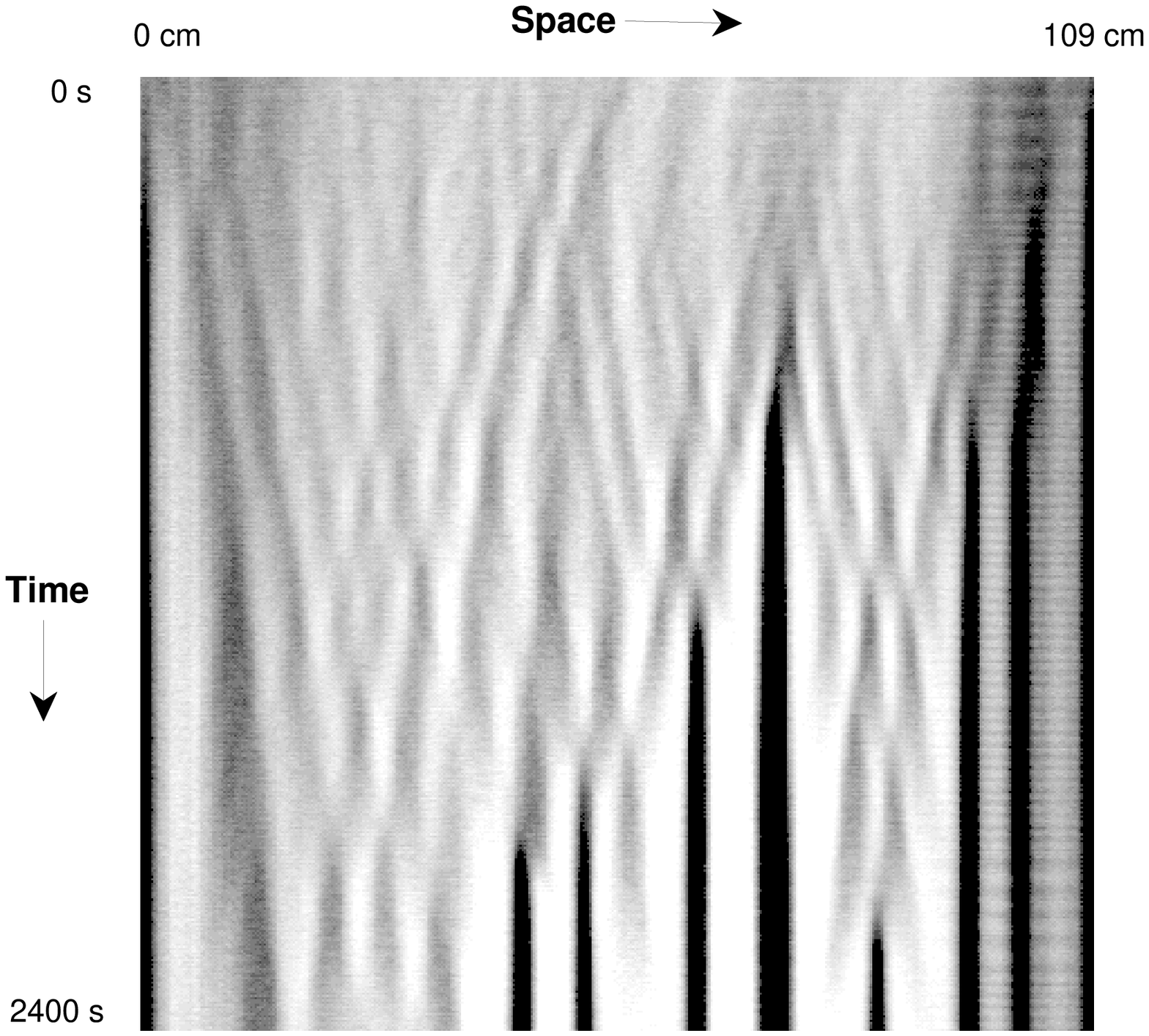}}
\epsfxsize =5in
\centerline{\epsffile{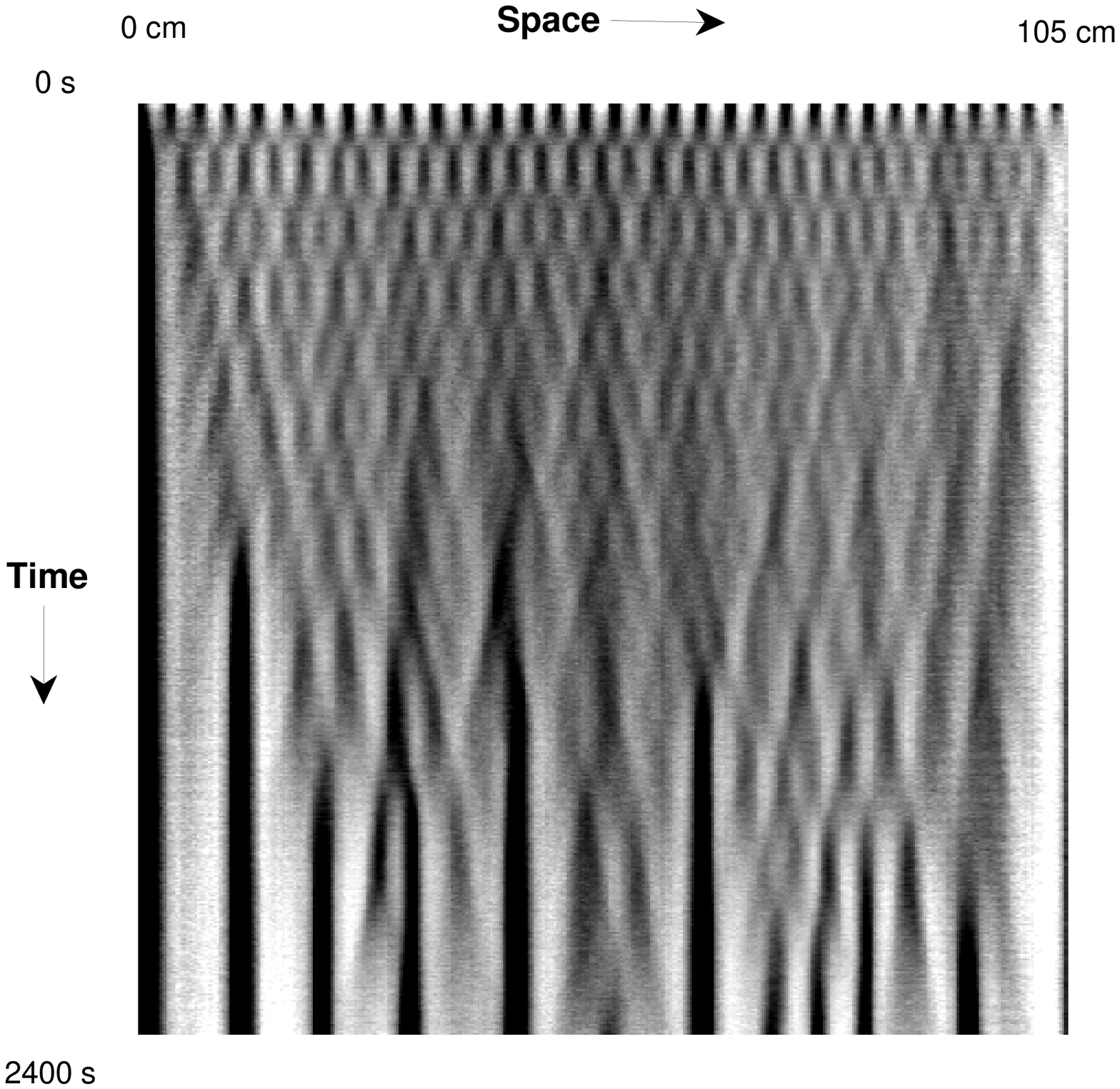}}
\vfill{}
\epsfxsize =5in
\centerline{\epsffile{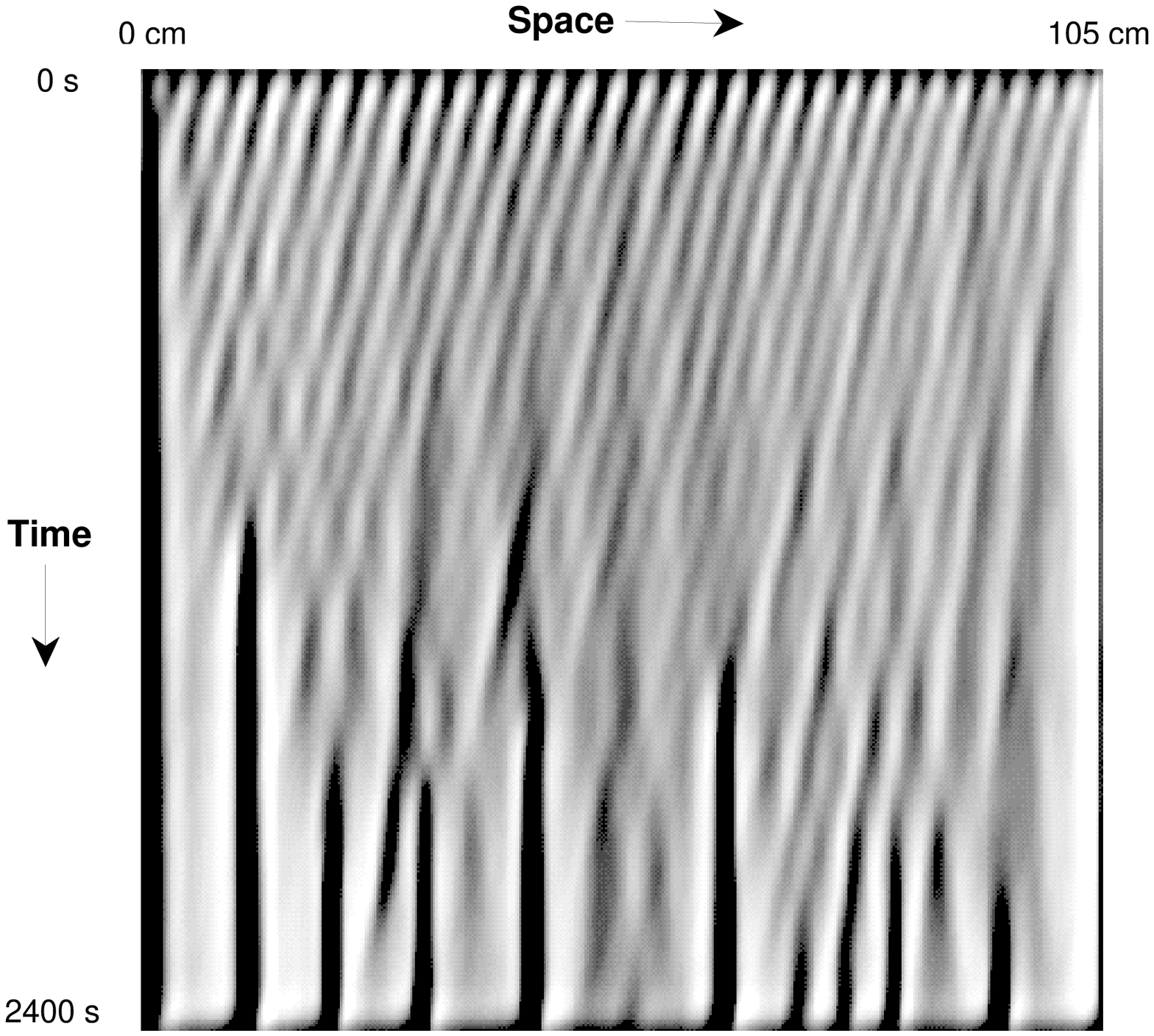}}
\vskip 1.0in
\caption{Space-time plots of the surface color.  Time runs downward.  Each frame shows the length of the tube and extends over 2400 $s$, or 1850 rotations.  (a) $\phi = 0.67$, premixed initial conditions. Patches of waves traveling in both directions evolve into larger stationary bands (thick vertical lines near the bottom).  (b) $\phi = 0.67$, periodically presegregated initial conditions with wavelength $3.2 \pm 0.1$ cm. Several cycles of a standing wave can be seen. (c) Same image as (b), but with the left-going waves extracted by a simple Fourier filter.}
\label{spacetime}
\vfill{}
\end{figure}

Qualitatively different behavior is observed when $\phi {{>}\atop{\sim}}0.55$.  Fig. \ref{spacetime}(a) shows the spacetime evolution of the pattern for $\phi = 0.67$. For premixed initial conditions, we find that patches of traveling waves evolve which propagate in both directions and have nearly equal wavelength and speed regardless of their position in the tube. These waves appear to pass through one another in the early stages, occasionally producing patches of standing waves, but later overlap to produce a region of spatiotemporal disorder.  From this disorder, large stationary bands emerge which are similar to those found at lower $\phi$. The stationary bands appear to be regions of nearly perfect segregation, while the traveling waves are much fainter and always contain a significant concentration of both sizes. The waves are not accompanied by a significant change in the dynamic streaming angle.  They have no obvious structure in the streaming direction and are equally visible from the back side of the tube in the non-streaming, packed portion of the rotation. The waves are always narrower than the final stationary bands and their wavelength and speed depends on $\phi$. Their motion is not related to any helical pattern; typical waves take several hundred tube rotations to travel one wavelength.

In order to investigate the traveling waves in a more controlled way, we employed presegregated initial conditions in the form of almost pure bands of sand and salt. The degree of segregation was therefore much larger than the initial fluctuations which we presume were amplified to produce the traveling waves in the premixed experiments. However, when measured after a brief transient, the wave speeds observed with presegregated initial conditions are reasonably consistent with those of the spontaneously emergent traveling waves of the same wavelength.   As shown in Fig. \ref{spacetime}(b), the main effect of preparing the initial conditions is to eliminate the spatial patchiness of the traveling waves.  In the early stages, several cycles of standing waves are observed, corresponding to the waves traveling in both directions.  As before, these give way to spatiotemporal disorder and eventually to stationary bands. The effect of a small ($ \approx 0.1 ^{o}$) tilt of the tube is to break the symmetry of the left- and right-going waves.  A wave moving toward the uphill end of the tube travels slower, while a downhill one speeds up.  We used this effect to accurately level the tube so that the two wave speeds were equal, to within error.  

\begin{figure}
\epsfxsize =5.5in
\centerline{\epsffile{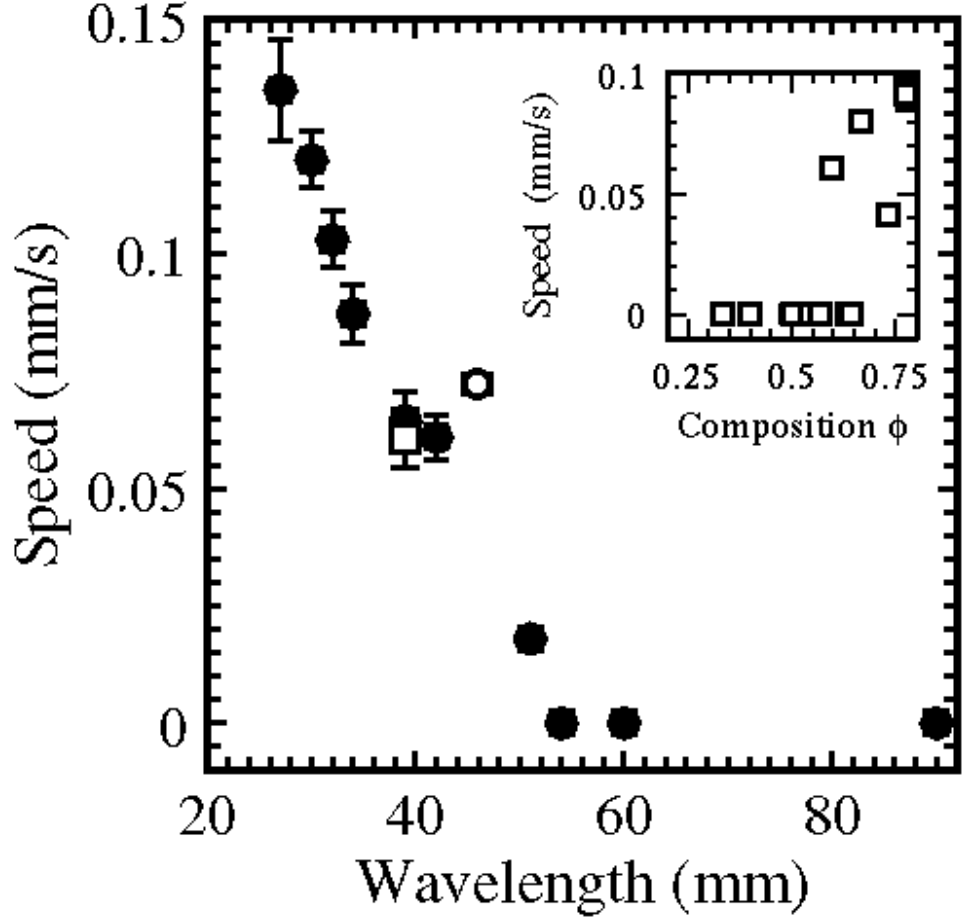}}
\caption{The wave speed {\it vs.} wavelength $\lambda$ for $\phi = 0.67$.  The solid symbols used presegregated initial conditions. The open symbols are from spontaneous waves using premixed initial conditions, with speeds found from Fourier peaks (open squares), and averages of slopes in spacetime images (open circles). For $\lambda {{>}\atop{\sim}} 5.4 \pm 0.1$ cm, the presegregated wavelength is stable. The inset shows the wave speed {\it vs.} $\phi$, using presegregated initial conditions with $\lambda = 3.2 \pm 0.1$ cm. }
\label{vlambda}
\end{figure}

It is straightforward to extract the left- and right-traveling waves by simple Fourier analysis of the space-time images so that the two wave components can be studied separately. An example is shown in Fig. \ref{spacetime}(c). The waves sometimes show a tendency to slow down as they evolve toward the disordered region.  The wave speed at early times can be found by locating the times when the standing waves pass through zero.  We used this to find the wave speed as a function of wavelength $\lambda$ for $\phi = 0.67$. To avoid early transients and later nonlinear effects, we used only the position of the first and second zero-crossing times.  The results are shown in Fig. \ref{vlambda}. For comparison, we also show speeds of some spontaneous waves occurring with premixed initial conditions. We find that the wave speed decreases with increasing $\lambda$, falling to zero for $\lambda {{>}\atop{\sim}} 5.40$ cm.  The cutoff presumably occurs when the presegregated black bands are wide enough that they remain pure even after rotational mixing redistributes some material axially.  Such pure bands are effectively impenetrable barriers to the movement of salt, and thus they cannot travel. Close to the cutoff, we find that some bands out of the presegregated initial configuration travel, while others do not.  This may be due to imperfections in the spacing of the initial configuration.  

A high composition fraction $\phi$ is required to observe the traveling waves, at least for this choice of grains.  The inset of Fig. \ref{vlambda} shows the wave speed as a function of $\phi$, using presegregated initial conditions.  For $\phi {{<}\atop{\sim}} 0.55$, no traveling waves are observed, only large bands.  Just below $\phi = 0.55$, we observe more early splitting and merging of the large bands than is observed at smaller $\phi$.  This behavior changes into traveling waves for larger $\phi$.  Thereafter, the wave speed increases with increasing $\phi$. 

Similar transient traveling waves are observed for large $\phi$ in this mixture  over a wide range of $\omega$ and filled volume fraction.  However, the effect is sensitive to the shape and size of the grains.  Traveling waves can be suppressed by changes in the size distribution of the individual components.  When rounded Ottawa sand of the same size is substituted for the salt, no traveling waves are observed.

The simplest models of axial segregation are inconsistent with traveling waves.  One-dimensional models\cite{zik} can be constructed from the continuity equation for mass concentration $c(z,t)$,
\begin{equation}
\partial_t c = -\partial_z J,
\label{continuity}
\end{equation} 
where $J(c,z,t)$ is the axial concentration current.  It is usually assumed that $J$ has the form
\begin{equation}
J = (\beta - D) \partial_z c,
\label{current}
\end{equation}
where $D>0$ is the usual Fick diffusion coefficient and $\beta(c,z)$ describes the tendency of the mixture to segregate. If $\beta(c,z)$ is a constant and $\beta > D$, Eqn. \ref{continuity} reduces to a diffusion equation with a negative diffusion coefficient, so that initial concentration fluctuations grow, in analogy with spinodal decomposition. Various assumptions for $\beta(c,z)$ define a class of nonlinear segregation models. In models where surface effects are proposed as the segregating mechanism, $\beta(c,z)$ is presumed to be related to the differences in the dynamic angles of repose of the two components. The most complete theory of axial segregation as a consequence of surface slope effects is that of Zik {\it et al.} \cite{zik} for  which $\beta(c) \propto c(1-c)$. 

All such models have the general feature that they are purely first order in time. This is not inconsistent with the appearance and irreversible merging of segregated bands. The traveling waves we observe are sufficiently linear at early times that they superpose to form standing waves to an excellent approximation.  This behavior makes it seem unlikely that they are the result of nonlinearity in a one-dimensional PDE model which is merely first order in time. No such equation can describe counter-propagating waves which are sufficiently linear that they obey superposition.  

Since we only imaged the streaming surface, it is conceivable that three-dimensional effects which involve the radially segregated core may be involved.  There exist no sufficiently large-scale simulations of three-dimensional segregation in drum mixers, nor any models of the interplay between the axial and radial modes of segregation.  Thus, we have limited insight into possible mechanisms for the travelling waves which might involve the core. It would be very interesting to use MRI\cite{mri} to examine the internal structure of traveling bands. Our present mixture would have no contrast under MRI, unfortunately, since neither component contains hydrogen.   

However the concentration is distributed, it is difficult to explain how any sort of wave-like dynamics of concentration could result from the highly dissipative underlying motions of the grains. The traveling waves do not appear to be associated with any collective motion of the grains.  Fauve {\it et al} \cite{fauve} observed an oscillatory cellular instability of the streaming surface and undulation modes of the lower contact line in experiments with homogeneous sand. These occurred on a timescale comparable to one rotation period.  So far as we can determine visually, neither of these motions are present in our experiment. In any case, the time scale of the traveling waves is much larger, suggesting that a different mechanism is involved.

Small electrostatic effects are more difficult to rule out.  Charge transferred between the grains or onto the tube could conceivably modify the segregation patterns by introducing long-range forces.\cite{electro}  Rubbing the exterior of the rotating tube or repeated rapid pouring of the mixture when it was outside the tube caused some of the grains to stick to glass surfaces. Given the highly insulating nature of all the components, charging effects are likely to be present at some level. No sticking or humidity-dependent effects were observed in normal operation, however.  Similar waves were observed using a plastic tube. Thus, we have no compelling candidates for the explanation of the origin of the waves, other than the random collisions of the grains themselves.

In conclusion, we have described a new phenomenon in a long rotating drum mixer: transient traveling waves of segregation.  The waves propagate equally well in either direction and appear spontaneously from initial concentration fluctuations for premixed initial conditions.  They can also be prepared and studied using presegregated initial conditions.  We have characterized the dependence of their speed on wavelength and composition.  Waves cannot easily be explained by theories based on one-dimensional continuity equations. It would be useful to look for waves in other sizes, shapes and types of grains, particularly in conductors and mixtures removed from $\phi = 0.5$. Many interesting avenues remain to be explored, including the nonlinear interactions of the waves and the propagation of pulses. These observations present an interesting challenge to our basic understanding of the segregation process.

We would like to thank Elaine Lau, Eamon McKernan, Holly Cummins and Michael Baker for experimental assistance and Zahir Daya and Wayne Tokaruk for useful insights.  This work was supported by The Natural Sciences and Engineering Research Council of Canada.

\end{document}